%% file: ms.tex
\documentclass{emulateapj}
\usepackage{natbib}

\slugcomment{{\sc Accepted to ApJ:} June 15, 2007}

\newcommand{\msun}{M$_\odot$}
\newcommand{\kms}{km s$^{-1}$}
\newcommand{\masyr}{mas yr$^{-1}$}

\shorttitle{Parallactic Distance to Orion with VLBA}
\shortauthors{Sandstrom et al.}

\begin{document}


\title{A Parallactic Distance of $389^{+24}_{-21}$ parsecs to the 
Orion Nebula Cluster from Very Long Baseline Array Observations}

\author{
Karin M. Sandstrom,
J. E. G. Peek,
Geoffrey C. Bower,
Alberto D. Bolatto and 
Richard L. Plambeck}

\affil{
Department of Astronomy and Radio Astronomy Laboratory, University of
California at Berkeley,
Berkeley, CA 94720}

\email{karin@astro.berkeley.edu}


\begin{abstract}

We determine the parallax and proper motion of the flaring,
non-thermal radio star GMR A, a member of the Orion Nebula Cluster,
using Very Long Baseline Array observations.  Based on the parallax, we
measure a distance of $389^{+24}_{-21}$ parsecs to the source.  Our
measurement places the Orion Nebula Cluster considerably closer than
the canonical distance of $480\pm 80$ parsecs determined by
\citet{genzel81}.  A change of this magnitude in distance lowers the
luminosities of the stars in the cluster by a factor of $\sim 1.5$.
We briefly discuss two effects of this change---an increase in the age
spread of the pre-main sequence stars and better agreement between the
zero-age main-sequence and the temperatures and luminosities of
massive stars.

\end{abstract}



\section{Introduction}
\label{intro}

The Orion Nebula is the nearest example of ongoing massive star
formation.  As such, it has an important place in our understanding of
this fundamental process.  Observations of luminosities, masses and
linear scales in the Orion Nebula all depend on the distance to the
cluster. For many years, the most often used distance measurement to
Orion has been $480 \pm 80$ parsecs, as determined by
\citet{genzel81}.  The Hipparcos mission was only able to marginally
detect the parallax of one star in the cluster, resulting in a
distance of $361^{+168}_{-87}$ pc.  Other techniques of estimating the
distance to the cluster provide values with large uncertainties and
systematic errors, but which are generally consistent within the
Genzel et al. range.

At the canonical distance of 480 parsecs, the annual parallax of
an object in the Orion Nebula would be about  2 milli-arcseconds.  In
order to measure this parallax to high precision, angular resolution
far in excess of 1 milli-arsecond is necessary.  The Very Long Baseline
Array (VLBA) can achieve this degree of angular resolution, and it
has thus been a very useful tool for determining parallaxes for
radio sources out to a few kpc (e.g.
\citet{brisken02,chatterjee04,hachisuka06,loinard05}).  

In order to obtain a precise measurement of the distance to Orion
using VLBI, the target source must be compact and persistent.  A
number of compact radio sources have been observed in Orion, first by
\citet{garay87} and later by \citet{felli93a} and others.  At least
ten of the sources detected by these authors were identified as
nonthermal radio stars because of their compactness (unresolved by the
VLA A-array) and variability \citep{felli93b}.  \citet{bower03}
observed one of these sources, GMR A, over the course of an extreme
millimeter flaring event, during which its 86 GHz flux density
increased by a factor of 5 over a few hours.  Follow-up observations
with the VLBA detected GMR A at 15 and 22 GHz with a compact size of
less than 1 milli-arcsecond.  We obtained five additional epochs of
VLBA observations spaced over the following year to monitor the
astrometric motion of the source.  We detect GMR A in four of those
five observations, indicating that this source was persistent enough
to allow measurement of its proper motion and parallax, thus
determining the distance to Orion in a model-independent manner.    

Unlike other published distance measurements to Orion, the parallactic
distance presented here is model-independent.  \citet{genzel81}
determined a distance of $480 \pm 80$ pc by modeling the proper
motions and radial velocities of H$_2$O masers in the BN/KL region
with an expanding, thick shell---a technique which required
assumptions about the geometry of the system of masers.  Most other
available distance measurements to Orion rely on stellar models in
some way.  Because the maser distance is independent of stellar
properties it has become the canonical distance to Orion.  Recently,
\citet{jeffries07} have found a distance of $392 \pm 32$ pc based on
the statistical properties of rotation in pre-main sequence stars.  In
determining this distance, \citet{jeffries07} assume a spectral type -
effective temperature scale for pre-main sequence stars and a random
distribution of spin axes. In addition, they also must accurately
identify stars which are still in the accretion phase of their
pre-main sequence lifetime.  Parallax provides a fundamental measure
of the distance, avoiding the systematic uncertainties associated with
these other techniques.

Accurate knowledge of the distance to the Orion Nebula Cluster is
important for the general understanding of star-formation in the
region.  Luminosities are proportional to distance squared
and the ages of pre-main sequence stars are typically determined by
comparing their temperatures and luminosities with theoretical models
(for example, \citet{palla99}).  Changes in the luminosity of the
stars translate directly into changes in their inferred ages, and the
age distribution of the pre-main sequence stars in the cluster is a
key component of any model which attempts to explain the mode of
star-formation in the region.

In Section~\ref{sec:data} we describe the VLBA observations and in
Section~\ref{sec:reduction} we discuss their reduction.  In particular
we discuss the use of a dual calibrator method developed by
\citet{fomalont05}.  Section~\ref{sec:analysis} covers the analysis of
our positions for the source and presents our best fit values for
the parallax and proper motion.  In Section~\ref{sec:discussion} we
discuss how this measurement will affect the study of the Orion Nebula
and its star-formation and compare our measurement with previous
results.

\section{Observations and Data Reduction}

\subsection{VLBA Observations}\label{sec:data}

The observations were carried out at a frequency of 15 GHz with all
available VLBA antennas.  There were six epochs of VLBA observations:
an initial observation of the source in January of 2003 in the wake of
its flaring event and five epochs evenly spaced over one year to
adequately sample the entire parallactic ellipse of GMR A.  

To determine the astrometric position of GMR A, we use
phase-referencing to the bright quasar J0541$-$0541 which is located 1.6
degrees to the southeast of the target.  Additionally, to account for
phase variation due to atmospheric gradients between J0541$-$0541 and
GMR A, we observe a secondary calibrator J0529$-$0519 which is 1.3
degrees northwest of GMR A and approximately colinear with GMR A and
J0541$-$0541.  The data reduction utilizing these two calibrators in
the phase-referencing is described in the following section. Each
observation, excluding January 2003, consisted of alternating 40
second integrations on the two calibrator sources and on GMR A for 6
hours.  The first epoch, January 2003, was part of a campaign to
understand the millimeter and x-ray flare from GMR A at high
resolution.  The secondary calibrator was not included in this track,
which consisted of alternating observations of only GMR A and
J0541$-$0541 for 40 seconds each. A very bright calibrator, J0530$+$1331,
was observed a few times during the course of each track for use in
removing instrumental phase offsets and delays in each IF.

\subsection{Data Reduction}\label{sec:reduction}

The data reduction was performed in AIPS.  The initial
calibrations proceeded as described in the AIPS cookbook for
VLBA data reduction.  These include amplitude calibration and
fringe-fitting to determine the instrumental single-band delay 
(slope of phase vs. frequency introduced by the instrument) and 
the residual phase delays and rates introduced by the atmosphere and
inaccuracies in the correlator model.

Recently discovered errors in the Earth Orientation Parameters (EOPS)
used by the VLBA correlator for all epochs excluding January 2003 were
corrected using the task CLCOR as described in the VLBA Memo
69\footnote{VLBA Test Memo 69, October 6, 2005
\url{http://www.vlba.nrao.edu/memos/test/test69memo/index.html}}.  The
data were corrected for the single- and multi-band delays with the
task FRING.  The target source is detected at 5-$\sigma$ or greater,
in four out of the six epochs using only phase-referencing to the main
calibrator J0541$-$0541.  In the October 2004 observation, GMR A is
only detected after ATMCA correction as described below.  In April
2004, there are a number of significant peaks in the image because the
flux from GMR A has been scattered by poor phase coherence, probably
due to fluctuations in tropospheric water vapor.  We have omitted the
April 2004 data from the following analysis because of its poor
quality.

In order to improve the image quality and remove some systematic
errors in the position of the target source we used dual-source
phase-referencing as implemented in the AIPS task ATMCA
\citep{fomalont05}.  This corrects for phase gradients across the sky
due to tropospheric effects.  The correction can be done in a number
of ways, but in our case we used interpolation between two calibrator
sources placed on either side of the target.  Prior to ATMCA we also
correct for the effects of structure in our calibrators through
self-calibration cycles on both sources.  It is clear from the images
that both calibrators have resolved structure, which makes the
self-calibration cycle important for accurate imaging of the target
source.  We note that the structure of the main calibrator appears to
be approximately constant over the time period of our observations. We
see at most a few percent variation in the brightness distribution of
J0541$-$0541 comparing between epochs with clean components all
restored to the lowest resolution of the five observations.  Because
of the lack of variation in the source structure and the high
signal-to-noise of the calibrator observations, the self-calibration
cycle should be sufficient to correct systematic errors in the phase
solutions due to the structure of the calibrator.  After ATMCA
calibration we see an increase in the compactness of GMR A of up to 17
percent.  Additionally, in the October dataset, the significance of
the GMR A detection changes from $<4$ to 5 sigma.  The effects of the
dual-source calibration are summarized in Table~\ref{tab:atmca}.

\input{tab1.tex}
\input{tab2.tex}
\input{tab3.tex}

\begin{figure*}[h]
\centering
\epsscale{1.0}
\plotone{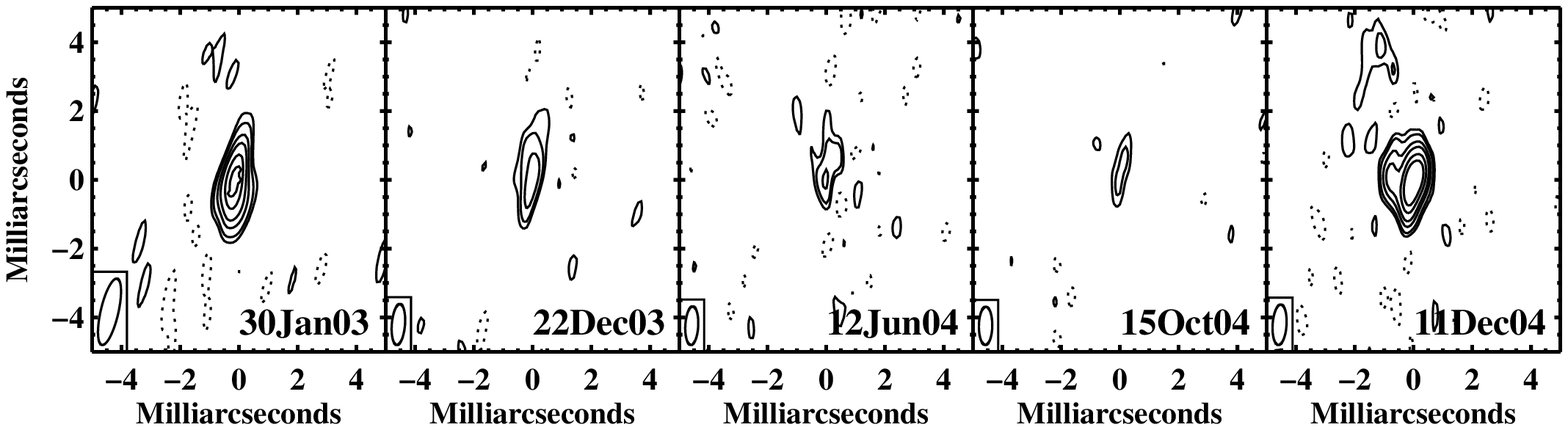}
\caption{Images of GMR A for all epochs.  All images are after ATMCA
calibration, except for the January 2003 image, as described in the
text. Each image is centered on the brightest pixel of that image.
The contour levels are -2.5, 2.5, 4, 7, 10, 15 and 20  times the noise
level of the individual image. The dotted contours are the negative
values.  The noise levels are listed in Table~\ref{tab:gmra}. The
sythesized beam is shown in the lower left corner.}
\label{fig:gmra}
\end{figure*}

\begin{figure*}
\centering
\epsscale{1.0}
\plotone{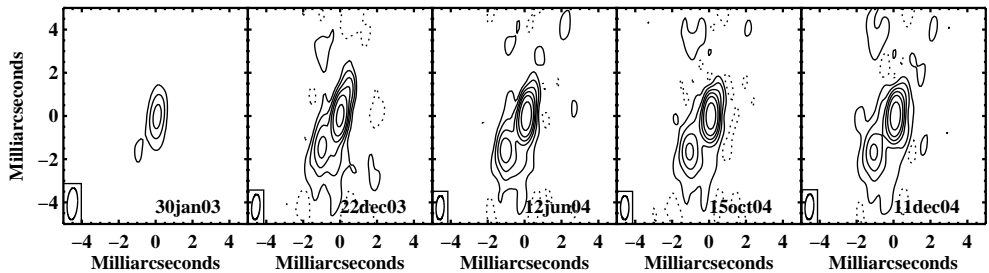}
\caption{Images of the primary calibrator at each epoch 
with ATMCA calibration, except for the January 2003 observation which
does not have ATMCA calibration but has been through an amplitude and
phase self-calibration cycle.  The images are centered on the location
of the brightest pixel in the December 2003 image.  The contour levels
are -5, 5, 20, 50, 100, 200, 500  times the noise level of the
individual image.  The synthesized beam is shown in the lower left of
each image. Negative contours are shown with dotted lines. In the
January 2003 image there are no pixels less than -5 times the noise
level.}
\label{fig:j0541}
\end{figure*}

\begin{figure*}
\centering
\epsscale{1.0}
\plotone{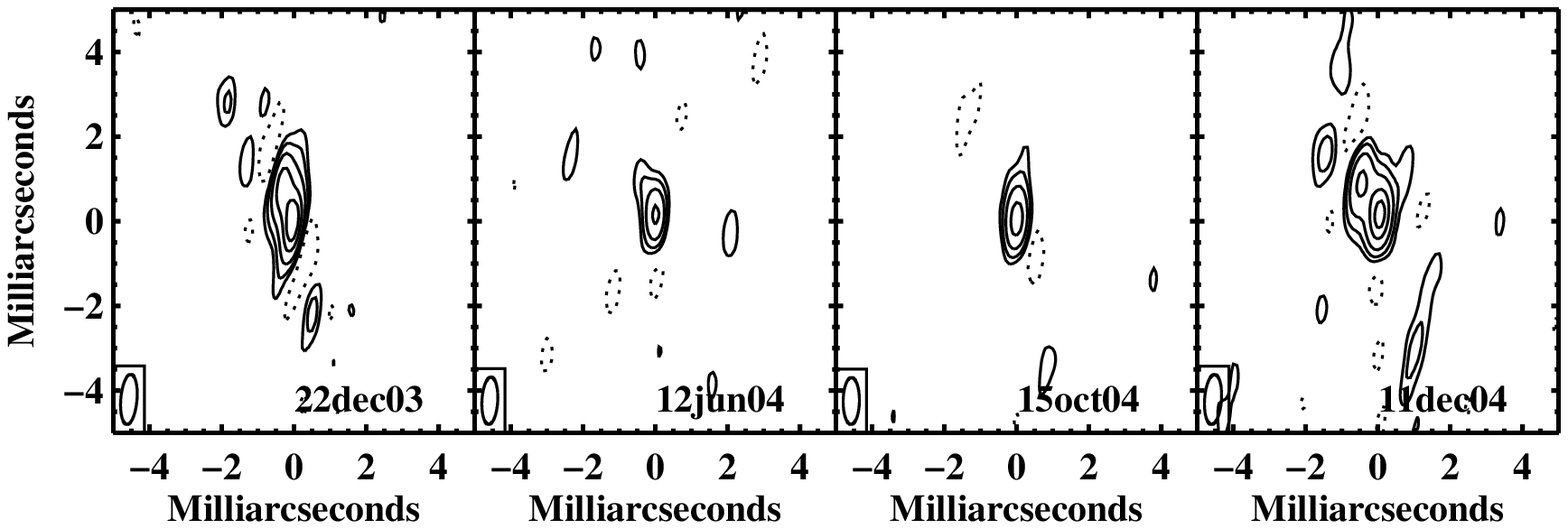}
\caption{Images of the secondary calibrator, with ATMCA calibration.
The images are centered on the location of the brightest pixel in the
December 2003 image.  The contours are -5, 5, 10, 20, 40 and 60 times
the noise level of the individual image, The synthesized beam is shown
in the lower left of each image.  Negative contours are shown 
with dotted lines.}
\label{fig:j0529}
\end{figure*}

After ATMCA, cleaned images of the sources were made using the task
IMAGR.  In each epoch, we measured the position of GMR A with
reference to the main calibration source J0541$-$0541 for which we
assume the J2000.00 position R.A.  $5^{\rm h}41^{\rm m}38\fs084106$,
Dec. $-5^\circ41\arcmin49\farcs42841$.  The position assumed for the
secondary calibrator, J0529$-$0519 in J2000.00 is R.A. $5^{\rm
h}29^{\rm m}53\fs532715$, Dec.  $-5^\circ19\arcmin41\farcs61564$.
This position is used in the course of ATMCA calibration.  The five
images of GMR A are shown in Figure~\ref{fig:gmra}, centered on the
maximum point in each image.  The images of the calibrators
J0541$-$0541 and J0529$-$0519 are shown in Figures~\ref{fig:j0541}
and~\ref{fig:j0529}, respectively.  In these figures the positions are
relative to the maximum of the image in the December 2003 observation.  

It is clear from the integrated flux densities listed in
Table~\ref{tab:gmra} that we observe significant variability in GMR A
at 15 GHz.  This observation is in line with previous studies that
show large changes in the source flux density
\citep{felli93b,zapata04,bower03}.  \citet{bower03} postulate that the
structure of GMR A consists of a compact, highly variable source that
the VLBA detects and an extended, $\sim 5$ mJy envelope that is
resolved out.  Our observations certainly indicate that the source
detected by the VLBA is highly variable. 

\subsubsection{Structure in the VLBI Images of GMR A}

The images of GMR A shown in Figure~\ref{fig:gmra} show resolved
structure.  At the distance we measure to the cluster, stellar
photospheres should be unresolved.  We can estimate the photospheric
radius of GMR A from the bolometric luminosity of 6 L$_\odot$ and the
spectral type K5 V as determined by \citet{bower03}, given the
effective temperature - spectral type calibration for pre-main
sequence stars of \citet{cohen79}.  This yields a radius of $\sim 4.3$
R$_\odot$, which at the distance of Orion is $\sim 50$ $\mu$as, well
below the resolution of our observations.  Since there is resolved
structure present in the images, we must consider its source in
determining the position of GMR A.  Scattering by turbulent
interstellar plasma can cause angular broadening.  We estimate the
effects of interstellar scattering on our images using the
\citet{cordes02} model, which predicts an angular broadening scale at
15 GHz of $\sim 3 \ \mu$as for the line of sight to the calibrators
and $\sim 0.05 \ \mu$as for GMR A.  The measured size of our
calibrators are consistent with very little or no interstellar
scattering, in agreement with the model predictions. Some of the
structure in our images is most likely due to atmospheric calibration
errors.  In particular, the correction from ATMCA applies mostly in
the Right Ascension dimension due to the positioning of our two
calibrator sources East-West relative to GMR A.  It is the case that
most of the images are elongated more in the Declination dimension
compared to the beam shape, suggesting atmospheric effects.  

However, there is also more persistent structure, mainly obvious when
GMR A is bright.  \citet{bower03} noted this structure and found that
the January 2003 observation could be equally well represented by a
Gaussian extended relative to the beam shape or by two unresolved
sources with a separation of 0.8 mas.  In the December 2003 and 2004
observations as well, we see structures that are better represented by
two unresolved sources or an extended Gaussian. We considered the
possibility of binarity to explain the resolved structure and we note
that if GMR A were a binary, orbital motion would introduce scatter
into our solution for the parallax and proper motion.  Using the
positions determined by fitting two unresolved sources to the images,
we attempted to find a mass ratio which gave a center of mass position
between the two purported sources that decreased the scatter in the
parallax and proper motion solution.  Because there is no consistent
way to distinguish between the two sources, we also searched through
the 16 possible permutations of primary and secondary components.  For
each of these permutations, we also attempted to fit a relative visual
binary orbit using a technique based on that of \citet{hartkopf89}.
We were unable to find a consistent solution that improved the scatter
of our parallax and proper motion using these two complementary
techniques.  Therefore, it seems unlikely that the structure in our
images is due to a binary companion. 

A more likely explanation for the structure in our images is large
scale magnetic features associated with the star.  These have been
observed with VLBI in other magnetically active weak-line T Tauri
stars \citep{phillips91,andre92} with sizes of
up to 10-20 R$_*$.  A separation of 0.8 mas at the distance we measure
to Orion corresponds to $\sim 15$ R$_*$, which is comparable to
structures observed on other weak-line T Tauri stars.  We do know that
GMR A has a $\sim 2$ kG photospheric magnetic field \citep{bower03}
and experiences a high level of magnetic flaring activity as seen in
x-ray, millimeter and radio observations, so it seems reasonable to
expect large scale magnetic structures around the star. The effect
of the positional jitter introduced in our measurements by imperfectly
calibrated atmosphere and by source structure is to increase the error
bars in the distance determination.

\subsubsection{Astrometric Positions of GMR A}

If GMR A were unresolved and we could account for the effects of the
atmosphere by observing a known point source, the best technique for
determining the positions of the target would be to fit a fixed-size
Gaussian to the images.  The size of the Gaussian would take into
account the blurring effects of the atmosphere.  However, since GMR A
appears to have resolved structure which may vary with time and
neither of our calibrators are unresolved, the best technique
for determining the source positions is to fit variable size
Gaussians to the images.  We did this using the AIPS task JMFIT,
allowing the dimensions and position angle of the fit Gaussian to
vary.  Table~\ref{tab:fits} lists the dimensions of the beam for each
observation and the properties of the best fit Gaussian.  The
positions derived from these fits are listed in Table~\ref{tab:gmra}.
As a check, we have also fit the images with a Gaussian constrained to
have the shape and size of the beam.  The positions we find from these
two techniques are consistent within their error bars, however the
positions we determine from fitting the constrained Gaussian have
smaller error bars, increasing the $\chi^2$ of our parallax and proper
motion solution, though it yields the same distance.  

\input{tab4.tex}

In the case of the January 2003 observation, ATMCA can not be applied,
as only one calibrator was observed.  To account for any systematic
differences in the positions determined with and without ATMCA, we
increased the uncertainty of the January 2003 observation by adding
the r.m.s. effect of ATMCA on the other three observations in
quadrature with its observed position errors (85 $\mu$as in R.A.  and
18 $\mu$as in Dec.).

\section{Astrometric Analysis}\label{sec:analysis}

The position of GMR A as a function of time is determined by its
position, parallax and proper motion in the following way:
\begin{equation}
\alpha(t) = \alpha_0 + \mu_{\alpha}t + \pi f_{\alpha}(\alpha,\delta,t)
\end{equation}
\begin{equation}
\delta(t) = \delta_0 + \mu_{\delta}t + \pi f_{\delta}(\alpha,\delta,t)
\end{equation}
Here $\mu_{\alpha}$ and $\mu_{\delta}$ are the proper motions in Right
Ascension and Declination, respectively, and $\pi$ is the parallax.
$f_{\alpha}$ and $f_{\delta}$ are the parallactic displacements for a
source at a distance of 1 pc at the position of GMR A.  The parallactic
displacements are calculated based on the formulae presented in the
U.S. Naval Observatory Almanac.

To fit to these five positions (ten data points) we employed a
$\chi^2$ fit to five parameters: position in R.A. and Dec, proper
motion in R.A.  and Dec., and parallax.  The best fit solution had a
reduced $\chi^2$ value of $\chi_R^2 = 10.2$. This indicates that our
error bars on the position are approximately $\chi_R = 3.2$ times
larger than the formal errors from Gaussian fitting, mostly due to
systematic effects from tropospheric variations and variability of the
source structure.  To determine the error bars on the best fit
parallax and proper motion we proceeded in the following way: first,
we added systematic errors in quadrature to the error bars listed in
Table~\ref{tab:gmra} to achieve $\chi_R^2 = 1$; we then did a Monte
Carlo simulation in which we added offsets, drawn from a Gaussian
distribution centered at zero with a width representing the total
positional errors from the previous step, to the measured positions
and fit for the astrometric parameters.  The resulting distributions
of parallax and proper motion were well-represented by Gaussians,
allowing a straightforward determination of the 1-$\sigma$ errors.  

The best approach to adding in systematic errors to achieve $\chi_R^2
= 1$ is not well-defined, so we approach the problem with a few
different techniques.  To start, we added in quadrature the same
systematic error in the R.A. and Dec dimensions till we reached
$\chi_R^2 = 1$.  This gave a parallax of $2.53 \pm 0.18$ mas
(equivalent to a distance of $395^{+30}_{-26}$ pc).  However, it is
most likely not the case that the systematic errors are the same in
R.A.  and Dec.  We know that our correction from ATMCA mostly applies
in the Right Ascension dimension, due to the positioning of our
calibrators East-West relative to the target.  To address the
non-uniformity of the systematic errors we added in quadrature error
ellipses with varying axial ratios and determined the axial ratio for
which the total area of the systematic error ellipse necessary to
achieve $\chi_R^2 = 1$ was smallest.  As expected based on the
positions of our calibrator sources, the necessary systematic error
ellipse is larger in the Declination dimension by a factor of 2.5.  To
reach $\chi_R^2 = 1$ the geometric mean of the axes of the error
ellipse with these dimensions was 0.17 mas.  This more realistic
appraisal of our systematic errors gives a parallax of $2.57 \pm 0.15$
mas ($389^{+24}_{-21}$ pc).  It is also possible that the systematic
errors are epoch dependent.  In this situation it is not clear how
best to add in systematic errors unless we base them on the beam shape
or the errors from Gaussian fitting (which are propotional to each
other in theory, $\sigma \sim 0.5
\frac{\theta_{\mathrm{FWHM}}}{\mathrm{SNR}}$).  To explore this
possibility we have scaled up the error ellipses from Gaussian fitting
to achieve $\chi_R^2 = 1$.  This technique gives a parallax of $2.61
\pm 0.14$ mas ($383^{+22}_{-20}$ pc), consistent with the other
techniques.

\begin{figure*}
\centering
\epsscale{0.9}
\plotone{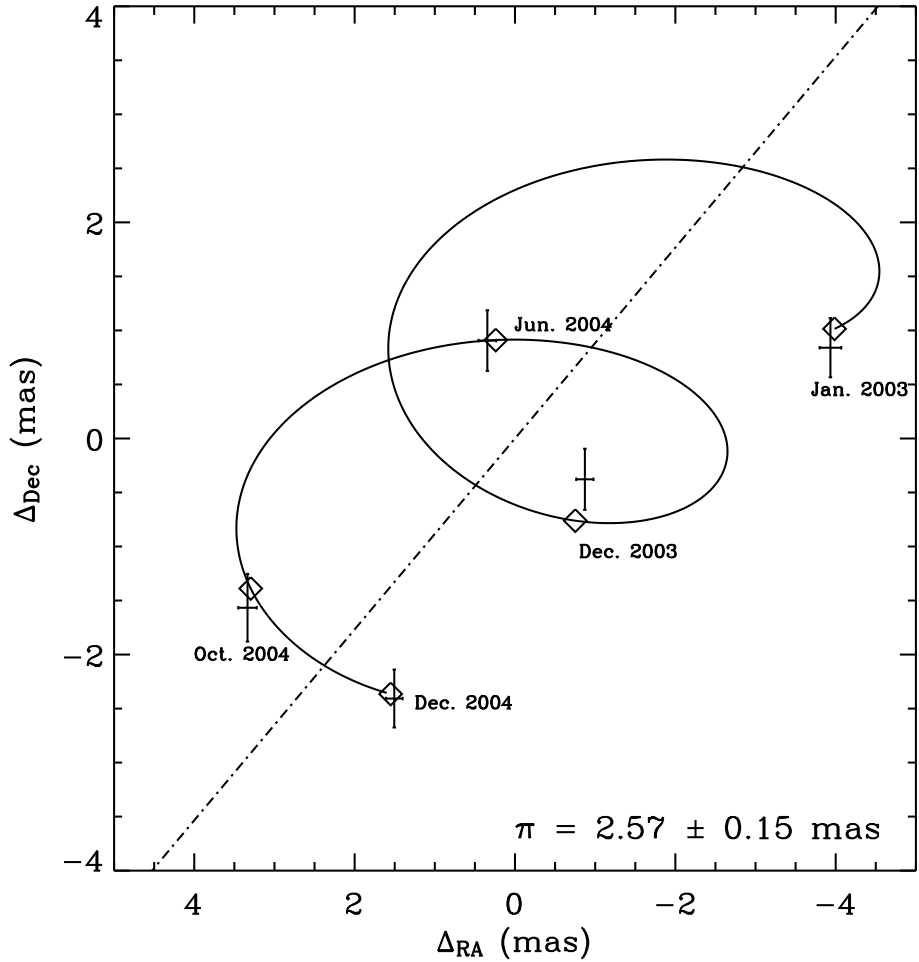}
\caption{The measured positions of GMR A with the best fit parallax
and proper motion.  The diamonds represent the predicted
position of GMR A for each observation.  The error bars on the
measured positions  are scaled as described in the text.  The dashed
line is the proper motion of the source, with the parallactic motion
subtracted. The parallax corresponds to a distance of
$389^{+24}_{-21}$ pc.}
\label{fig:solution}
\end{figure*}

We note that the astrometric parameters obtained through these various
techniques are robust to our treatment of the errors.  We consider
adding the fixed axial ratio error ellipse to be the most realistic
appraisal of the systematic errors given the positioning of the
calibrators.  This technique yields a parallax of $\pi = 2.57 \pm
0.15$ mas, which corresponds to a distance of $389^{+24}_{-21}$ pc.
The proper motions of GMR A in R.A. and Dec. from this solution are
$\mu_{\alpha} \cos{\delta} = 1.89 \pm 0.12$ \masyr \ and $\mu_\delta =
-1.67 \pm 0.19 $ \masyr.  At the distance we measure, the transverse
velocity is $v_t = 4.65 \pm 0.39$ \kms.  Figure~\ref{fig:solution}
shows a plot of our solution and Table~\ref{tab:params} lists the
measured astrometric parameters and their uncertainties.

\input{tab5.tex}

\section{Discussion}\label{sec:discussion}

The Orion Nebula is part of a very large and very complex
star-formation region (see \citet{genzel89} for a review of the large
scale structure).  The implications of a measurement of the distance
to one star thus depend sensitively on where that star is located
relative to the stellar associations and molecular gas in this region.
The available evidence very strongly constrains GMR A to be a part of
the Orion Nebula Cluster, located within a few parsecs of the
Trapezium stars but embedded in the molecular cloud which is currently
being disrupted by the ionizing radiation from $\Theta^1$ Ori C.  In
the following paragraphs we will present the evidence for placing GMR
A in the Orion Nebula Cluster and then proceed to compare our distance
to previous measurements and, finally, briefly discuss some of the
implications our distance measurement has for the general study of
this important region.

\subsection{The Membership of GMR A in the Orion Nebula Cluster}

\citet{bower03} observed GMR A, following its intense outburst
detected with the BIMA interferometer, in the near-IR with NIRC and
NIRCSPEC on the Keck telescopes.  While there is no optical source
coincident with the position of GMR A, there is an infrared source
which they identify as a weak-line T Tauri star embedded in molecular
gas.  GMR A was also detected in the COUP survey as a highly variable
x-ray source (COUP J053551.8$-$052149) \citep{feigelson02,getman03}
behind a gas column density of N$_{\mathrm{H}} = 10^{22.3}$ cm$^{-2}$.
The combination of the spectral classification as a K5 V star and a
bolometric luminosity of 6 L$_\odot$ \citep{bower03} suggests a very
young age for this star (around 1 Myr, based on the pre-main sequence
tracks of \citet{palla99})  as does its location embedded in molecular
gas.   

The molecular gas in Orion A, located behind the H II region created
by the Trapezium stars, has a very high column depth $A_V \approx
50-100$ mag, essentially providing a wall behind which no background
sources are detected.  Near-IR surveys have shown that the projected
spatial distribution of optically visible and extincted, near-IR
sources are very similar \citep{hillenbrand98} and that around 50\% of
sources are only visible in the infrared.  The three-dimensional
distribution of the Orion Nebula Cluster has thus been interpreted as
fairly spherical with about half of that sphere still embedded in the
molecular cloud out of which it formed \citep{hillenbrand98}.  The
projected distribution of stars has a radius of $\sim 3$ pc around
$\Theta^1$ Ori C.  GMR A is located 1.95 arcminutes away from the
center of the cluster, which corresponds to a projected distance of
0.22 parsecs at our measured distance to the cluster.  The x-ray
absorption column density measured to GMR A indicates that it must be
embedded in the molecular gas, but it cannot be more than a few
parsecs deep.  The lack of foreground and background star-forming
regions, the proximity of GMR A to the Trapezium on the sky, the fact
that it is embedded in molecular gas, and its very young age
convincingly place GMR A as a member of the Orion Nebula Cluster.  

In addressing the membership of a stellar cluster, a typical technique
is to compare the velocity of a star to the average velocity and
dispersion of the cluster.  There have been a number of measurements
of these quantities for the ONC with proper motion and radial velocity
studies.  Recent spectrocopic observations of stars in the cluster by
\citet{sicilia-aguilar05} show a mean heliocentric velocity of 25
\kms, with a $\sim 2$ \kms \ uncertainty in the zero point and a
dispersion of $\sigma = 2.3$ \kms.  \citet{bower03} measured a
heliocentric radial velocity of $14\pm 5$ \kms \ for GMR A.  Given the
combined uncertainties in these measurements and the cluster radial
velocity dispersion, GMR A is consistent with the radial velocity of
the cluster.  

Most proper motion studies of clusters, including that of
\citet{jones88} which surveyed nearly 1000 stars in Orion, measure
relative proper motions.  In order to compare our measurement, which
is an absolute proper motion, to the results of these studies, we must
first find the absolute proper motion of the ONC to subtract from our
values.  This value has been measured by a number of authors using
different astrometric catalogs.  \citet{baumgardt00} use the Hipparcos
catalog and measure a mean proper motion of $\mu_\alpha \cos(\delta) =
1.73 \pm 0.40$ \masyr \ and $\mu_\delta = -0.47 \pm 0.27$ \masyr \ for
the ONC.  \citet{kharchenko03} used the ASCC catalog (which contains
data from Hipparocs, Tycho and the USNO catalogs) and found very
similar values $\mu_\alpha \cos(\delta) = 2.02 \pm 0.74$ \masyr \ and
$\mu_\delta = -0.19 \pm 0.66$ \masyr.  Later, \citet{kharchenko05}
improved their determination of the absolute proper motion with an
expanded astrometric catalog and found $\mu_\alpha \cos(\delta) = 1.96
\pm 0.31$ and $\mu_\delta = -0.77 \pm 0.46$ \masyr.  All of these
values are consistent within their error bars, so we proceed in our
analysis using the recent value of \citet{kharchenko05}.  We measure a
proper motion for GMR A of $\mu_\alpha \cos(\delta) = 1.89 \pm 0.12$
\masyr \ and  $\mu_\delta = -1.67 \pm 0.19$ \masyr.  These values are
quite similar to the cluster mean.  We can compare the residual
velocity after subtracting the mean to the velocity dispersion of the
cluster.  The velocity dispersion of the ONC has been measured by a
number of authors.  The largest survey of relative proper motions was
carried out by \citet{jones88}.  They found a one-dimensional proper
motion dispersion of $\sim 1.1$ \masyr \ for all of the cluster stars
in their survey (I magnitude of 16 or higher).  Additionally, they
found a trend of decreasing velocity dispersion with increasing mass
(and therefore, magnitude) for stars in the ONC.
\citet{hillenbrand98} show that for stars with masses between 0.1 and
0.3 \msun \ the dispersion is $<\sigma> \approx 1.26$ \masyr \ and for
stars between 1 and 3 \msun \ it decreases to $1.00$ \masyr.  The
proper motion of GMR A is less than 1-$\sigma$ from the cluster
average indicating a very high membership probability.

At the distance we measure to the ONC, a velocity of 1 \masyr \ is
equivalent to 1.85 \kms.  Relative to the cluster, GMR A is moving
$0.13 \pm 0.61$ \kms \ West, $1.67 \pm 0.93$ \kms \ South, and $-11.0 \pm
5.4$ \kms \ along the line of sight.  We note that there are large
uncertainties in the radial velocity, both for GMR A and for the
cluster as a whole.  However, even at $-11$ \kms \ relative to the
cluster, GMR A would only have moved $\sim 10$ pc over its lifetime.

There is one measurement of the proper motion of GMR A in the
literature by \citet{gomez05}.  They determined the absolute
proper motions of 35 radio sources in Orion with archival VLA data
spanning 15 years.  Although we have much higher angular resolution in
our observations, their long time baseline makes the comparison
useful.  They measure the absolute proper motion of the cluster to be
$\mu_\alpha \cos(\delta) = 0.8 \pm 0.2$ \masyr \ and $\mu_\delta
\cos(\delta) = -2.3 \pm 0.2$ \masyr.  These values are quite different
from others in the literature.  The absolute proper motion for GMR A
they find is $\mu_\alpha \cos(\delta) = -2.02 \pm 1.87$ \masyr \ and
$\mu_\delta = 1.15 \pm 1.93$ \masyr.  Our measurement is only 
consistent with these values at the 2-sigma level. They
find a substantially higher velocity dispersion for the cluster than
found in optical surveys and theorize that the population of radio
sources might have larger random velocities.  However, using our
proper motion, GMR A has a very typical velocity compared to the
optical sources.

To summarize, GMR A is a very likely member of the Orion Nebula Cluster 
based on its youth, proximity to the cluster on the sky, proper
motion, and location embedded in molecular gas.  Having established
its place in the Orion region, we now compare our distance measurement
to those in the literature.

\subsection{Comparison with Previous Measurements} 

Despite its importance in our understanding of star-formation, the
distance to the Orion Nebula Cluster is quite uncertain.  Till
recently there has only been one model-independent distance
measurement---the marginially significant parallax from Hipparcos of
the star HD 37061 which corresponds to a distance of
$361^{+168}_{-87}$ parsecs \citep{bertout99}.  

Many authors have estimated the distance to the cluster by fits to the
upper main sequence, a procedure which is complicated by systematic
uncertainties in the models, the variable background and extinction of
the Orion Nebula, and the region over which stars are included in the
analysis, among other issues.  Some results with this technique are
those of \citet{penston73} who found a distance of $363^{+26}_{-24}$
parsecs, using infrared photometry to characterize each star's
extinction individually; a later, more detailed analysis by
\citet{penston75} who found $\sim 400 \pm 20$; \citet{warren78} found
a distance of $483^{+57}_{-51}$ parsecs; and \citet{anthony-twarog82}
who reanalyzed the data from \citet{warren78} found a distance of
$434^{+21}_{-19}$ parsecs.  These results alone---some of which make
use of the same data, others which use the same techniques---show a
spread of more than 100 parsecs, illustrating the difficulty of such
measurements.  Many authors have also attempted statistical techniques
which combine the proper motion and radial velocity distributions of
the stars with assumptions regarding cluster expansion or contraction.
Two early examples of this technique are \citet{strand58} who find a
distance of 525 parsecs and \citet{johnson65} who find 380 parsecs.  

Although the Hipparcos mission could only marginally detect the
parallax of a single star in the ONC, combination of the astrometric
measurements of many stars has yielded some results for Orion.
\citet{wilson05} analyzed the aggregate astrometric motions of stars
foreground and background to the Orion A in three distinct regions of
the cloud.  They estimate a distance of $465^{+75}_{-57}$ parsecs to
the ONC region.  This distance is larger than what we measure, but is
consistent with our measurement at the one-sigma level.
\citet{brown94} also used arguments regarding foreground and
background stars, but instead comparing their reddening and 100 $\mu$m
fluxes to estimate distances to the near and far edges of the cloud of
320 and 500 parsecs.

The orbital parameters of binary systems can be used in some cases to
determine very accurate distances.  \citet{stassun04} have analysed an
eclipsing binary 0.3 degrees south of $\Theta^1$ C Ori and found it to
be at $419 \pm 21$ parsecs.  However, its status as a member of the
ONC is somewhat uncertain because of its relatively old age and
location relative to the cluster.  The distance to this binary
agrees with our distance to the ONC.  Recent work by
\citet{kraus07} on the visual binary system containing $\Theta^1$Ori C
yields two equally good orbital solutions which their data cannot yet
distinguish.  Assuming a luminosity-mass relationship for the stars
they find distances of $384 \pm 11$ and $434 \pm 12$ parsecs from the
two orbits.  Further astrometric and spectroscopic observations of
this system may provide a very accurate distance in the future.

\citet{genzel81} combined observations of the proper motions and
radial velocites of H$_2$O masers in the BN/KL region with the
assumption of a spherical, uniformly expanding, thick shell and found
a distance of $480 \pm 80$ parsecs.  Because of its relative precision
and independence from stellar evolution calculations, this distance
has become the canonical distance to the Orion Nebula Cluster.
Indeed, many studies of the ONC population and star-formation make use
of this distance without consideration of its substantial
uncertainties.  Our distance measurement is 20\% closer than that
of \citet{genzel81}, though within their combined uncertainties.
However, our distance is more precise and does not depend on the
assumption of geometries for expanding maser sources which has been
shown to be considerably more complex in many cases than what these
authors assume \citep{greenhill05}.  

Very recently, \citet{hirota07} measured the annual parallax of one H$_2$O
maser spot in the Orion BN/KL outflow using VERA (VLBI Exploration of
Radio Astronomy).  The parallax they measure corresponds to a distance
of $437 \pm 19$ parsecs using only the motion in Right Ascension, and
$445 \pm 42$ when they solve for the parallax using both the Right
Ascension and Declination. Note that these error bars are only
statistical, and the latter value agrees with ours at the 1-$\sigma$
level.  Parallax determinations from maser spot motion can be
problematic, because of intrinsic structure changes and/or spot
acceleration.  For the spot in question \citeauthor{hirota07} detect
changes in its emission line profile over their two years of
observation, which likely indicate changes in the structure of the
source.  Stellar parallax measurements are less vulnerable to this
type of systematic uncertainty.  If the difference between our VLBA
measurement and the \citeauthor{hirota07} VERA determination is real,
and the VERA error bars are accurate, it may argue for a larger
separation between the BN/KL region and the Orion Nebula Cluster along
the line of sight.

Finally, we compare our distance to the recent work of
\citet{jeffries07} who have used the statistical properites of
pre-main sequence star rotation to determine a distance of $392 \pm
32$ parsecs to the ONC.  This techniques involves the assumption of
random spin orientations for the stars in the cluster and makes use of a
spectral type - effective temperature scale for pre-main sequence
stars.  From the whole sample analyzed by \citet{jeffries07}, a
distance of $440 \pm 34$ parsecs is derived.  However, excluding stars
which show evidence of accretion---a factor which makes the necessary
radius determination unreliable---lowers the distance to $392 \pm 32$
parsecs.  Our distance agrees very well with this latter value.
\citet{jeffries07} note that the distance they determine is very
sensitive to the assumed spectral type - effective temperature scale,
such that distance scales with the square of the effective
temperature.  In addition, it is clear that the exclusion of Classical
T Tauri stars, which show evidence for accretion, significantly
changes the derived distance, indicating that an accurate account of the
various sources of systematic error is extremely important to this
technique.  

To summarize, we present a distance to the ONC of $389^{+24}_{-21}$
parsecs using the fundamental technique of parallax.  In contrast to
previous distance estimates, this measurement does not rely on
modeling of stellar evolution, cluster dynamics, or calibrations of
effective temperature for pre-main sequence stars.  In comparing our
measurement to previous values, we find it to be consistent with many
prior measurements within their substantial error bars.  We find good
agreement with recent values from \citet{jeffries07}, \citet{kraus07}
and \citet{stassun04}, but independent of the various assumptions
those authors are forced to make.

\subsection{Implications of a Closer Distance to the ONC}

Many studies of the Orion Nebula assume the distance of
\citet{genzel81} or similar.  Our distance is 20\%  
closer. In luminosity terms, this change means luminosities are a
factor of 1.5 lower than previously claimed, a result that is
especially important for the determination of pre-main sequence star
ages, which scale with luminosity as $t\propto L^{-3/2}$.  Therefore,
at a distance of 390 parsecs, the stars are nearly twice as old as
they would be at 480 parsecs. This scaling of age with luminosity is
true only for fully convective pre-main sequence stars that have
contracted by a substantial amount from their initial radii (see, for
instance, \citet{palla99} for further discussion).  Near the birthline
this assumption breaks down and the ages will not be affected to the
same degree.  A decrease in luminosity will therefore age the entire
population of pre-main sequence stars, but not uniformly, increasing
the age spread of the population.

The distribution of stellar ages in Orion is the basis for many
theories describing star-formation in the region and massive,
clustered star-formation in general.  These theories can be broken
down into two general categories---those that purport that star
formation in the ONC happened suddenly and quickly
(e.g. \citet{hartmann01,hartmann03}) and those that argue that star
formation has been occuring over a longer timescale, but has been
accelerating in recent times (e.g.  \citet{palla00,tan06,huff06}).
The age spread of cluster members is a fundamental measurement of the
timescale of star-formation in the cluster.  A larger age spread, as
would result from the decrease in luminosity, tends to favor models
where star-formation has occurred over a more extended period of time.

Another interesting feature of the stellar population in Orion is that
the high mass stars seem to fall above the expected zero-age main
sequence for the cluster \citep{palla99,hillenbrand98}.
\citet{palla99} argue that this must be due to a systematic problem in
determining the luminosities, effective temperatures or some
combination thereof.  These analyses assumed a distance of 470
parsecs, so there is indeed a systematic offset---according to our
measurement these luminosities are too high by a factor of 1.5.  A
shift of this magnitude brings the luminosities of the high mass stars
into much better agreement with the zero-age main sequence
predictions.

\section{Conclusions}\label{sec:conclusions}

We have monitored the astrometric motion of the flaring, non-thermal
radio star GMR A over the course of two years with the Very Long
Baseline Array.  We determine from these data the proper motion and
parallax of the star.  Based on its young age, its proximity to the
center of the Trapezium on the sky, the consistency of its proper
motion with that of the cluster and its location embedded in molecular
gas, the probability that GMR A is a member of the Orion Nebula
Cluster is very high, and thus the distance we determine based on its
parallax is representative of the cluster as a whole.  We find the ONC
is at a distance of $389^{+24}_{-21}$ parsecs, nearly 100 parsecs
closer than the canonical distance of 480 parsecs determined by
\citet{genzel81}.  The distance presented here is in good agreement
with recent work by \citet{jeffries07}, with the advantage of being
independent of assumptions about stellar properties.  A closer
distance has important implications for the study of star-formation in
the Orion Nebula, one of the most well-studied sites of massive star
formation, most notably the increase in the ages and the age spread of
the pre-main sequence stars in the cluster.  The decrease in
luminosity also brings the more massive stars into better agreement
with the zero-age main-sequence.

Further VLBA observations of other radio stars in the ONC may overcome
the limitations of a single star distance and begin to probe the depth
of the cluster.  In addition, suitable targets for this type of
observation should be found in most clusters with substantial pre-main
sequence populations.  Future VLBI observations of magnetically
active,  pre-main sequence stars in these clusters could provide
precise, fundamental distance measurements to many nearby star-forming
regions.

\acknowledgements

The authors would like to thank the referee for thorough and helpful
comments.  We would also like to thank Steve Stahler and Reinhard
Genzel for sharing their expertise on the Orion region.  KMS
acknowledges support from an NSF Graduate Research Fellowship and
would like to thank Franck Marchis and Jason Wright for helpful
discussions. The research of JEGP was is supported in part by NSF
grant AST04-06987. The National Radio Astronomy Observatory is a
facility of the National Science Foundation operated under cooperative
agreement by Associated Universities, Inc.

{\it Facilities:} \facility{VLBA ()}



\clearpage

\end{document}

%% file: tab1.tex
\begin{deluxetable*}{lcccccc}
\tablewidth{0pt}
\tabletypesize{\scriptsize}
\tablecolumns{7}
\tablecaption{GMR A Positions and Fluxes}
\tablehead{ 
\multicolumn{1}{l}{Date} &
\multicolumn{1}{c}{R.A.} &
\multicolumn{1}{c}{R.A. Error} &
\multicolumn{1}{c}{Dec.} &
\multicolumn{1}{c}{Dec. Error} & 
\multicolumn{1}{c}{Noise Level} &
\multicolumn{1}{c}{Int. Flux Density} \\
\colhead{} &
\colhead{(J2000)} &
\colhead{($\mu$as)} &
\colhead{(J2000)} &
\colhead{($\mu$as)} &
\multicolumn{1}{c}{(mJy/beam)} &
\multicolumn{1}{c}{(mJy)} } 
\startdata
29 Jan 2003 & 5 35 11.80269059 & 14.84 &  $-$5 21 49.246612 &  36.48 & 0.395 & 11.366 $\pm$ 0.785 \\
22 Dec 2003 & 5 35 11.80289569 & 29.89 &  $-$5 21 49.247830 &  93.57 & 0.267 & 6.815 $\pm$ 0.875 \\
12 Jun 2004 & 5 35 11.80297711 & 36.82 &  $-$5 21 49.246546 &  86.22 & 0.280 & 4.217 $\pm$ 0.743 \\
15 Oct 2004 & 5 35 11.80317743 & 50.88 &  $-$5 21 49.249019 & 163.29 & 0.271 & 2.791 $\pm$ 0.731 \\
11 Dec 2004 & 5 35 11.80305485 & 11.47 &  $-$5 21 49.249859 &  21.66 & 0.321 & 23.002 $\pm$ 0.974 
\enddata
\tablecomments{Positions measured relative to the assumed J2000.00 position
of J0541$-$0541, R.A.  $5^{\rm h}41^{\rm m}38\fs084106$, Dec. 
$-5^\circ41\arcmin49\farcs42841$. The errors listed are the formal
errors from Gaussian fitting.}
\label{tab:gmra}
\end{deluxetable*}

%% file: tab2.tex
\begin{deluxetable*}{lcccccc}
\tablewidth{0pt}
\tabletypesize{\scriptsize}
\tablecolumns{7}
\tablecaption{J0529$-$0519 Positions and Fluxes}
\tablehead{ 
\multicolumn{1}{l}{Date} &
\multicolumn{1}{c}{R.A.} &
\multicolumn{1}{c}{R.A. Error} &
\multicolumn{1}{c}{Dec.} &
\multicolumn{1}{c}{Dec. Error} & 
\multicolumn{1}{c}{Noise Level} &
\multicolumn{1}{c}{Int. Flux Density} \\
\multicolumn{1}{c}{} &
\multicolumn{1}{c}{(J2000)} &
\multicolumn{1}{c}{($\mu$as)} &
\multicolumn{1}{c}{(J2000)} &
\multicolumn{1}{c}{($\mu$as)} &
\multicolumn{1}{c}{(mJy/beam)} &
\multicolumn{1}{c}{(mJy)} }
\startdata
29 Jan 2003 & \nodata          & \nodata &  \nodata           &  \nodata & \nodata & \nodata          \\
22 Dec 2003 & 5 29 53.53352197 & 2.867   & $-$5 19 41.616801  &  8.036   & 2.042   & 178.80$\pm$ 3.09 \\
12 Jun 2004 & 5 29 53.53351859 & 4.541   & $-$5 19 41.616825  & 10.928   & 2.415   & 120.91$\pm$ 4.60 \\
15 Oct 2004 & 5 29 53.53351577 & 3.804   & $-$5 19 41.616909  &  9.789   & 2.863   & 142.64$\pm$ 4.27 \\
11 Dec 2004 & 5 29 53.53351540 & 4.153   & $-$5 19 41.616692  &  8.757   & 2.399   & 174.45$\pm$ 3.70 
\enddata
\label{tab:j0529}
\tablecomments{Positions measured relative to the assumed J2000.00
position of J0541$-$0541, R.A.  $5^{\rm h}41^{\rm m}38\fs084106$, Dec. 
$-5^\circ41\arcmin49\farcs42841$.  The errors listed are the formal
errors from Gaussian fitting.}
\end{deluxetable*}

%% file: tab3.tex
\begin{deluxetable*}{lccccc}
\tablewidth{0pt}
\tabletypesize{\scriptsize}
\tablecolumns{6}
\tablecaption{Effects of ATMCA Calibration}
\tablehead{ 
\multicolumn{1}{l}{Date} &
\multicolumn{1}{c}{$\Delta_{\mathrm{R.A.}}$} &
\multicolumn{1}{c}{$\Delta_{\mathrm{Dec}}$} &
\multicolumn{1}{c}{Total Shift} &
\multicolumn{1}{c}{\% Change Area\tablenotemark{a}} & 
\multicolumn{1}{c}{\% Change Peak Flux\tablenotemark{b}} \\
\colhead{} &
\colhead{($\mu$as)} &
\colhead{($\mu$as)} &
\colhead{($\mu$as)} &
\colhead{} &
\colhead{} }
\startdata
29 Jan 2003 & \nodata  & \nodata  & \nodata  & \nodata & \nodata \\
22 Dec 2003 & $-4.36$  & $115.92$ & $116.02$ & $-$13.2 & $+$20.6 \\
12 Jun 2004 & $-84.60$ & $88.92$  & $122.46$ & $-$17.4 & $-$15.9 \\
15 Oct 2004 & \nodata  & \nodata  & \nodata  & \nodata & \nodata \\
11 Dec 2004 & $+85.32$ & $83.16$  & $119.14$ & $-$11.5 & $+$24.1 
\enddata
\tablenotetext{a}{The area of the source is defined here as
$\pi a b$ where the $a$ and $b$ values are the
FWHM of the major and minor axes of the best fit Gaussian.}
\tablenotetext{b}{The peak flux is measured in units of Jy/beam.}
\label{tab:atmca}
\end{deluxetable*}

%% file: tab4.tex
\begin{deluxetable*}{lcccc}
\tablewidth{0pt}
\tabletypesize{\scriptsize}
\tablecolumns{5}
\tablecaption{Beam Sizes and Best-Fit Gaussian Dimensions}
\tablehead{ 
\multicolumn{1}{l}{Date} &
\multicolumn{1}{c}{Beam FWHM} &
\multicolumn{1}{c}{Beam P. A.} &
\multicolumn{1}{c}{Gaussian FWHM} &
\multicolumn{1}{c}{Gaussian P. A.} \\
\multicolumn{1}{l}{} &
\multicolumn{1}{c}{(mas $\times$ mas)} &
\multicolumn{1}{c}{($^{\circ}$)} &
\multicolumn{1}{c}{(mas $\times$ mas)} &
\multicolumn{1}{c}{($^{\circ}$)} }
\startdata
29 Jan 2003 & 1.994 $\times$ 0.597 & -15.52 & 2.028 $\times$ 0.738 & -10.00 \\
22 Dec 2003 & 1.195 $\times$ 0.448 &  -5.58 & 2.230 $\times$ 0.619 & -9.52  \\
12 Jun 2004 & 1.122 $\times$ 0.438 &  -3.63 & 1.682 $\times$ 0.576 & -14.13 \\
15 Oct 2004 & 1.111 $\times$ 0.455 &  -1.04 & 2.114 $\times$ 0.482 & -11.81 \\
11 Dec 2004 & 1.178 $\times$ 0.473 &  -4.53 & 1.662 $\times$ 0.800 & -12.18  
\enddata
\label{tab:fits}
\end{deluxetable*}

%% file: tab5.tex
\begin{deluxetable*}{ll}
\tablewidth{0pt}
\tabletypesize{\scriptsize}
\tablecolumns{2}
\tablecaption{GMR A Astrometric Parameters\tablenotemark{*}}
\tablehead{ 
\multicolumn{1}{l}{Parameter} &
\multicolumn{1}{l}{Value} } 
\startdata
\cutinhead{Fit Quantities}
Epoch & 2004.25 (MJD 53095.3)  \\
Right Ascension $\alpha_0$ & $5^{\rm h} 35^{\rm m} 11\fs80295404$ \\
Declination $\delta_0$ & $-5^\circ 21\arcmin 49\farcs247452$  \\
$\mu_\alpha \cos{\delta}$ (\masyr) & 1.89 $\pm$ 0.12 \\    
$\mu_\delta$ (\masyr) & $-$1.67 $\pm$ 0.19 \\
Parallax $\pi$ (mas) & 2.57 $\pm$ 0.15 \\
\cutinhead{Derived Quantites}
Distance (pc) & $389^{+24}_{-21}$ \\
Transverse Velocity (\kms) & 4.65 $\pm$ 0.39 \\ 
Position Angle\tablenotemark{\dag} ($^\circ$) & 131.5 $\pm$ 3.7
\enddata
\label{tab:params}
\tablenotetext{*}{All coordinates are listed in the J2000 equinox and
measured in reference to the assumed position of J0541$-$0541: 
$5^{\rm h}41^{\rm m}38\fs084106$, Dec. $-5^\circ41\arcmin49\farcs42841$ .}
\tablenotetext{\dag}{The position angle of the proper motion, measured 
from North through East.}
\end{deluxetable*}